\documentclass[12pt,twoside]{article}
\usepackage{fleqn,espcrc1,epsfig}

\title{Valence pairing, core deformation and the development of
two-neutron halos}

\author{F.M.~Nunes\address{ NSCL and Department of Physics and Astronomy
Michigan State University, East Lansing MI 48824 USA}
\thanks{This work has been partially
supported by  the NSCL at Michigan State University}}

\begin{document}

\maketitle

\begin{abstract}
We explore the evolution of the structure of the ground state of a nucleus
with two valence nucleons as the system approaches the two particle
threshold. We use a three-body model of $core+n+n$
where the core is deformed and allowed to excite. We find that both
NN correlations and correlations due to deformation/excitation of the core
inhibit the formation of halos. Our results suggest that it is unlikely
to find halo nuclei on the dripline of deformed nuclei.
\end{abstract}

\section{Motivation}

One of the most interesting results from experiments with
Radioactive Beams was the discovery of the halo phenomenon
\cite{dima,thompson}. Nuclear Halos can develop when the system
approaches a threshold and the relative motion is not constrained
by a strong long range repulsive force. Then the tail of the wavefunction
extends out well beyond the region of the nuclear interaction,
generating an unusual outer region of low nuclear density. In the
well known examples of $^6$He or $^{11}$Li the valence neutrons
spend a large fraction of their time in an s- and/or p-wave motion
relative to the core. In such systems, the neutrons are well
decoupled from the core, which is often taken to be inert. Three
body models \cite{he6,li11} have been very successful in
describing a variety of properties of these Borromean systems. The
decoupling of valence nucleons and core-nucleon degrees of freedom
is not a good approximation when we move away from the dripline.
In most cases, somewhere between the valley of stability and the
dripline, the valence nucleons are somewhat correlated with the
nucleons in the core\footnote{If the valence neutron and the core
are completely uncorrelated, the system's wavefunction can be
written as a single product of the core's wavefunction and the
valence particles' wavefunction. In general this is not the
case.}. An effective way to take this correlation into account
without solving the many body problem is providing the core with
collective degrees of freedom. These could in principle be derived
microscopically but can also be introduced phenomenologically. An
extension of the three-body model (core+N+N) to include core
excitation and deformation was performed and applied to $^{11}$Be,
$^{12}$Be and $^{14}$Be
\cite{be11nunes,be12nunes1,be12nunes2,be14}.

The observation of nuclear halos has been limited to light
nuclei on the driplines (mainly the neutron dripline, although a
couple of cases have been found on the proton dripline).
Toward heavier nuclei the neutron dripline
is not well defined. One very interesting question is whether
halos exist for heavier systems. This question has been
partly addressed within Hartree-Fock-Bogolyubov (HFB) looking specifically
 into the issue of pairing \cite{antihalo}. Results show that,
for a given isotope, by adding neutrons, pairing prevents the
divergence of the rms radius and thus hinders the appearance of
the halo state. More recently similar studies \cite{pair1} show
that,  as the valence neutron binding energy artificially
approaches zero, the contribution of the pair correlation is very
low for single-particle s-waves, suggesting the appearance of the halo. 
Further work \cite{pair2} shows explicitly that the halo phenomenon
appears even in the presence of stronger many-body pair
correlations. In any case, one should keep in mind that the 
single particle s-wave valence neutron is so decoupled from 
the mean field when approaching
threshold that a HFB description may not adequate.

In heavier systems one often needs to consider deformation. The
possibility of a deformed one-neutron halo \cite{misu} was studied
within a Nilsson type model, but using a spheroidal basis. Results
in \cite{misu} show that even with deformation, one-neutron halo
states completely decoupled from the rest of the system can appear
in the limit of low binding. Recent work \cite{def1} has looked
into the effect of deformation within the single particle  Nilsson
model, without pairing. It is shown that the s-wave component
becomes dominant as the binding energy of the system approaches
zero, irrespective of deformation. In other words, deformation
does not hinder the formation of halos. The p-wave orbitals were
also studied in detail \cite{def2} although they are less likely
to produce halo states. Note that in both \cite{def1,def2} only
the one-neutron halo case was considered. Here, we are interested
in two-neutron halos.

The few-body models for halos \cite{he6} take into account the few
body dynamics between the valence-halo nucleons and the core
exactly, whilst oversimplifying the interaction with the core. The
decoupling approximation of core and halo degrees of freedom is
valid for low binding energy, consequently one can expect that the
few-body models with core excitation are the adequate tool to
explore the possibility of existence of halos in intermediate mass
nuclei. The effect of deformation can be studied in a natural way,
within the deformed core model developed in Ref.
\cite{be12nunes1}. Pairing in the sense discussed in \cite{pair1}
does not have an easy translation into the few-body nomenclature.
Some microscopic pairing is already included effectively through
the phenomenological core-n interaction. The only pairing
explicitly taken into account is that of the valence NN
correlation through the S-wave component of the NN interaction. We
will come back to this point at a later stage.

With the aim of exploring the possibility of two-neutron halo
states in heavier nuclei, we look at the effects of pairing and
deformation on the formation of halo states, using a three-body
model with core deformation/excitation. In section 2. we briefly
introduce the model, definitions, and some technical details. 
In section 3. the results are shown and
compared with previous findings. Finally conclusions are drawn in
section 4.

\section{Technical considerations and some definitions}

It is clear that a one-neutron halo state is most likely to appear
if the occupancy of  s-wave components is large, given that the
centrifugal barrier will hinder the appearance of the tail. Thus a
large occupancy of the $l=0$ orbit, or at most $l=1$, as the binding
energy tends to zero, is a necessary condition for the appearance
of the halo phenomenon (consistent with the divergence of the rms
radius \cite{radius}). When there are two valence neutrons outside
the core, the situation for a halo is not as straightforward and
this will be the focus of the present work. We will not discuss
the proton halo case, as then the Coulomb barrier further hinders
its development.

\begin{figure}[b!]
\vspace{-1cm}
\centerline{
    \parbox[t]{0.8\textwidth}{
\centerline{\psfig{figure=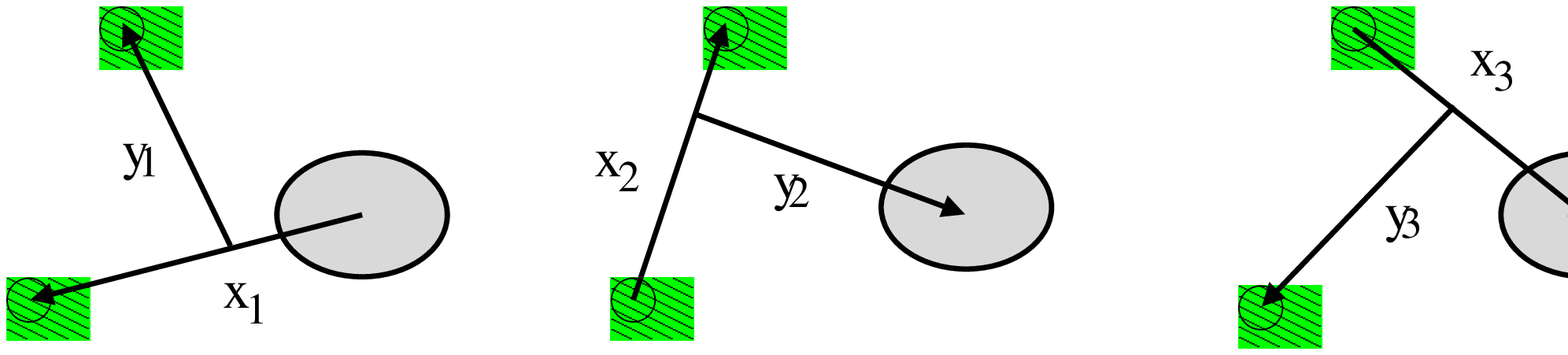,width=0.8\textwidth}}
\vspace{-1cm}
    \caption{Jacobi coordinates for the three-body problem. }
\label{jacobi}}
}
\end{figure}

Next we  present a few technical
considerations to introduce the adopted
condition for the appearance of the two-neutron halo phenomenon.
For the description of two-neutron halos,
it is usual to express the three-body
system in Jacobi coordinates ($x_i,y_i$),
represented in Fig.{\ref{jacobi}} (where $i=1,2,3$ refers to a
particular Jacobi set).
The Jacobi coordinates can be transformed into the
hyperspherical coordinates: the hyper-radius
$\rho^2  =  x_i^2 +y_i^2 = \sum_{i}^{3} A_i r_i^2$
related to the size of the three-body system, and
the hyper-angle $\theta_i  =  \arctan(\frac{x_i}{y_i})$
related to the correlations between the two Jacobi variables.
The hyperspherical expansion represents
the three-body wavefunction in a particular $i$ Jacobi coordinate set,
in terms of known polynomials containing the
angular and hyper-angular dependence (see \cite{be12nunes1} for more details).
If $\psi^{i,J} (x_i, y_i)$ is the three-body wavefunction written in
the $i$ Jacobi coordinates, and $(l_{xi}, l_{yi})$ are the associated
orbital angular momenta, then:
\begin{eqnarray}
\psi^{i,J} (x_i,y_i)& = & \rho_i^{-\frac{5}{2}}
\sum_{K_i} \; \chi^{i,J}_{\alpha_i K_i } (\rho) \;\;
\varphi_{K_i}^{l_{xi} l_{yi}}(\theta_i)\;\;,
\label{eq:hwf1}\\
\mbox{with} \;\;\;\;\;
\varphi_{K_i}^{l_{xi} l_{yi}}(\theta_i) & = & N_{K_i}^{l_{xi} l_{yi}} \;
(\sin\theta)^{l_{xi}}  \;  (\cos\theta)^{l_{yi}} \;\;
P_{n_i}^{l_{xi} + 1/2, l_{yi} + 1/2}(\cos2\theta_i) \;\;.
\label{eq:hwf2}
\end{eqnarray}
The Jacobi polynomial $ P_{ni}^{l_{xi} + 1/2, l_{yi} + 1/2}$,
normalized by $N_{K_i}^{l_{xi} l_{yi}}$, depends on
a new quantum number, $K_i$, the \mbox{{\it hyper-momentum}}.
$K_i$ is directly related to the order of the corresponding
Jacobi polynomial $K_i=l_{xi}+l_{yi}+2n_i$ ($n_i$=0,1,2,...).
All other quantum numbers are represented by $\alpha_i$,
including internal spins and relative orbital angular momenta.

The introduction of the hyperspherical coordinates and the
hyperspherical expansion mentioned above is extremely useful,
since it  reduces the  three-body problem to coupled hyper-radial
equations of the form \cite{face}:
\begin{equation}
(-\frac{\hbar^2}{2m} \frac{d^2}{d\rho^2}+
\frac{\hbar^2 (K_i+3/2)(K_i+5/2)}{(2m\rho^2)}-E)\chi^i_{\alpha_iK_i}(\rho) +
\sum_{j\alpha_jK_j}V^{ij}_{\alpha_iK_i,\alpha_jK_j}(\rho)
           \chi^j_{\alpha_jK_j}(\rho) = 0\;,
\label{coup-eq}
\end{equation}
where $m$ is an arbitrary mass unit.
For a detailed definition of the coupling potential
$V^{ij}_{\alpha_iK_i,\alpha_jK_j}(\rho)$ see \cite{face}.
Eq. \ref{coup-eq} shows that the centrifugal barrier
for two valence neutrons depends on the hyper-momentum,
which in turn relates
to the sum of the relative angular momenta between the three bodies.
For two-neutron halos to occur this centrifugal barrier needs to
be minimal, i.e. the occupancy of the K=0 component needs to be
large.  In \cite{dima}, it is stated that only the K=0 or K=1 components 
can give rise to a halo structure. For the ground state of this
system, there are no K=1 components. Thus, for our test case, 
the necessary condition
for the appearance of a two-neutron halo is that the K=0 component
should be larger than 50\%. This will be the criterion followed 
in our work. Some are more familiar with the use of a divergent radius
as a halo signature. It should be stressed that in the three-body case,
K=0 is the only component that produces a divergent rms radius.
Thus, the two conditions are closely related.

The hyperspherical transformation can be performed for any number
$N_v$ of valence neutrons, and the resulting coupled hyper-radial
equation will contain a centrifugal barrier similar to the one in
Eq. \ref{coup-eq}, but incremented. For this reason it has been
remarked before that one does not expect to find halos with
$N_v>2$. The conclusions to be drawn from our work are thus an
upper limit to any system with a larger number of valence
neutrons.

In the model derived in \cite{be12nunes1}, excitation appears
through the assumption of a rotational model for the core, with
a deformation $\beta$. The effective interaction
between the core-n is not central and contains higher multipoles
which couple different core states. The strength of those
couplings obviously depends on the deformation parameter $\beta$
which can be estimated from electric transition data between the
core's ground state and the relevant excited state. The particular
example studied in \cite{be12nunes1} takes into account the strong E2
transition between the core $0^+$ ground state and the $2^+$
excited state, through a quadrupole deformation $\beta_2$. 
Calculations can be easily generalized to
include any other transition within a collective model (for
example in \cite{na21} the model for $^{16}$O core also included
octupole deformation $\beta_3$ based on the large E3 connecting
the ground state $0^+$ with the first excited $3^-$ state).

In the three-body core+n+n model one can define several relative
orbital angular momenta. To avoid confusion, we will always use
lower case s, p, d, etc for the core-nucleon relative angular
momentum and capital letters S, P, D for the NN partial waves.

As mentioned before, pairing is partly embedded in the effective
core-n interaction. Relating mean field pairing with the NN
correlations seen in few-body problems is by no means trivial.
In some sense, the mean field pairing has more than the few-body
NN correlations we include. While in mean field pairing, all pairs
contribute, including those within the core, in the few-body case,
only the valence pair contributes explicitly, and any other
pairing contribution appears effectively in the fitted core-N
interaction. On the other hand, pairing in BCS is by definition
the correlation energy associated with the existence of
the bound cooper pair in the medium. In HFB it is
clearly separated from the total mean field. It is standard practice to
parameterize it as a delta function in the S-wave NN
channel, and only very recently have finite range effects 
been included \cite{frhfb}. In few-body models, the free NN 
interaction is included explicitly: it is finite range, L-dependent, 
with a tensor part,
such that it reproduces the low energy NN phase shifts. What we
can try to assess within the three-body model, is the importance
of the correlation between the two valence neutrons, by switching
off different parts of the NN interaction. It is clear that we
need to go beyond the few-body formalism to make accurate
predictions for heavy dripline nuclei, but we are still learning
how to generate halo phenomena in a microscopic mean field type
model. It is thus important to try to make the link between the
mean field and the few-body languages.

\section{Results}

\subsection{Fixed deformation}

Starting with the three-body model for $^{12}$Be \cite{be12nunes1}
we performed a series
of calculations to explore the possible development of the two-neutron halo when
the system is forced artificially to approach threshold.
We allow the core to exist in its ground state and its first
$2^+$ excited state at $E_x=3.368$ MeV.
In \cite{be12nunes1} the effective n-core interaction
corresponding to $\beta_2=0.67$ is modeled with a Woods-Saxon
plus spin-orbit term:
\begin{equation}
V^{be12}_{n-core}(\vec r)={\bf V}_{ws}(\vec r) +
(\vec l \cdot \vec s) \: {\bf V}_{so}(r)\, .
\label{eq:potential}
\end{equation}
The Woods-Saxon term depends on the orientation/excitation of the core,
and has the standard form:
\begin{equation}
V_{ws}(r,\theta,\phi)  =  \frac{V_{ws}}{1+e^{\left(\frac{r-R(\theta,\phi)}
{a_{ws}}\right) }},\:\:\: \:\:\:\:\:   R(\theta,\phi)  =  R_{ws}
( 1 + \beta\: Y_{20} (\theta,\phi) ).
\label{eq:ws}
\end{equation}
The spin-orbit term is undeformed and defined as
\begin{equation}
(\vec l \cdot \vec s) \: V_{so}(r)  =
-\:{(\frac{\hbar}{m_{\pi}c})}^2 (2 \vec l \cdot \vec s)
\frac{V_{so}}{r}\:\:\frac{d}{dr} \left[1+e^{\left(\frac{r-R_{so}}{a_{so}}
\right)} \right]^{-1}\:.
\label{eq:so}
\end{equation}
The Woods-Saxon depth is parity dependent
$V_{ws}(l=0,2)=-54.239$ MeV and $V_{ws}(l=1)=-49.672$ MeV, the radius is
$R_{ws}=2.4883$ fm and the diffuseness is $a_{ws}=0.65$ fm.
The spin-orbit has the same radius and diffuseness
as the Woods-Saxon part, and its strength is $V_{so}=-34.0$ MeV.

In this work, we force the system to move toward threshold by
artificially decreasing the strength of the interaction
$V_{n-core}(\vec r) = \lambda \; V^{be12}_{n-core}(\vec r)$.
Smaller values of  $\lambda$ will force '$^{11}$Be' to cross
threshold,  which in turn will provide smaller binding energies
for the '$^{12}$Be' three-body system. We explore the behaviour of
the ground state wavefunction of '$^{12}$Be' as its binding energy
tends to zero. Note that, in our procedure, the spin-orbit force
is scaled too. Throughout this work, we take the NN interaction
between the two valence neutrons to be the GPT interaction
\cite{gpt} (the same as in \cite{be12nunes1,be12nunes2}). This
interaction, built from the sum of three gaussians, contains
central S-, P- and D-terms, as well as a spin-orbit and a tensor
force.

Calculations for the ground state $0^+$
were performed in a truncated subspace and
checked for convergence. The main concern had to do with $K_{max}$,
the maximum value of hyper-momentum included in expansion (\ref{eq:hwf1}).
The smaller the two-neutron binding energy of the '$^{12}$Be' system,
the larger the required $K_{max}$.  As seen
before \cite{be12nunes1}, the probabilities associated
with the most significant parts of the wavefunctions converge
faster than the binding energy. Due to
computational limitations, we could only obtain convergence
for states with $S_{2n}\ge 0.08$ MeV, where $S_{2n}$ is
the two-neutron binding energy.

\begin{figure}[t!]
\vspace{-1cm}
\centerline{
    \parbox[t]{0.45\textwidth}{
\centerline{\psfig{figure=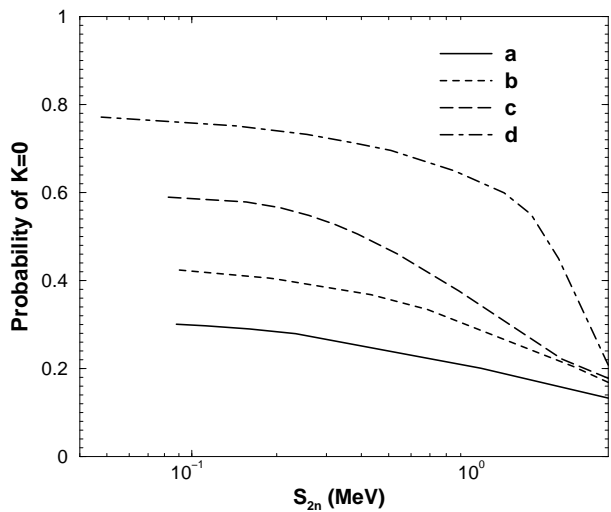,width=0.5\textwidth}}
\vspace{-1cm}
    \caption{Probability of the K=0 component in the
    ground state of a three-body core+n+n system based on
    $^{12}$Be: a) including both the NN interaction
    and core deformation; b) including only core deformation;
    c) including only the NN interaction; d) switching both
    the NN interaction and the core deformation to zero. }
\label{fig1}}
}
\end{figure}
Fig. \ref{fig1} shows the fraction of the ground state three-body
wavefunction that exists in the K=0 state. We consider four
distinct situations: a) the realistic case  where both the
deformation and the  NN interaction are included (solid line); b)
considering core deformation but switching off the NN interaction
(dotted line); c) including the NN interaction but keeping
$\beta_2=0$ (dashed line); and d) the most simple case, where
there is no NN interaction and no core excitation/reorientation
(dot-dashed line). For the $\beta_2=0$ results, the starting point
for the core-n interaction was refitted to obtain realistic
$1/2^+$ and $1/2^-$ energies for the $^{11}$Be system.

 Expectedly, in d), as the system approaches threshold, it
favours the K=0 component such that the probability of finding K=0
tends to $\lim_{S_{2n}\rightarrow 0}P(K=0) \approx 0.8$.
Introducing
the correlation between the two valence neutrons
reduces this to $\lim_{S_{2n}\rightarrow 0}P(K=0) \approx 0.6$.
According to the criterion in \cite{dima}, in both cases the
K=0 component is larger than $50$\% and thus the system
would be considered a halo in the low binding limit.
However, when core excitation comes into the picture,
the K=0 component saturates at much lower values
$\lim_{S_{2n}\rightarrow 0}P(K=0) \approx 0.4$
suggesting that in this situation no halo would develop.
When we take both NN GPT interaction and deformation,
$\lim_{S_{2n}\rightarrow 0}P(K=0) \approx 0.3$
When introducing deformation, the excitation (or reorientation)
coupling  mixes in higher
angular momentum components (in this case mainly d-waves)
reducing the probability of generating long halo tails
in the wavefunction.
When the two neutrons are strongly bound to the core, the core+N+N model
is not a good approximation, consequently we do not discuss the results for
$S_{2n}$ larger than $\approx 4$ MeV.
We  looked explicitly at the sum of all core excited components
as the binding approaches zero and found that these remain finite
($ > 20$ \%).

All four curves in Fig. \ref{fig1} have a similar dependence
on the binding energy.  We find that the K=0 occupancy
near threshold can be parameterized by the
three parameter polynomial expression
\begin{equation}
Prob(K=0)=\frac{P1}{(P2+log(S_{2n}))^{P3}}.
\end{equation}

We also checked the separate contributions of the NN interaction.
As mentioned before, the GPT interaction \cite{gpt} contains a
central term for S-, P- and  D-waves, a spin-orbit term and a tensor part.
Compared to results where the NN interaction was switched off,
it is mainly the tensor part that produces the
reduction from $d \rightarrow c$ observed in Fig. \ref{fig1}.

We now come back to the issue of translating these results into
the pairing language used in HFB \cite{pair1,pair2}, and other
mean field approaches.
Since we found that the tensor part of the NN interaction is the main
contributor to the hindrance of the halo,
the mean field pairing force, which contains S-waves only, would not
be able to reproduce the same effect. This issue needs to be
carefully addressed in future work.

Often, the dineutron model is used to describe reactions of
systems with two loosely bound neutrons. If there is a strong
correlation between the two valence neutrons, such that a
dineutron could be formed, then a halo in the two body sense (i.e.
$l_{y2}=0$ from Fig. 1) could appear: {\it the dineutron halo}
(this is most likely the case of $^6$He where $\approx 85$\%
of the wavefunction is in the $l_{y2}=0$). We have compared the
$l_{y2}=0$ components of the ground state wavefunction when
including the GPT NN interaction with that resulting from
switching off the NN interaction. The GPT interaction enhances the
$l_{y2}=0$ component by $\approx 3$\% and does not change the
energy behaviour significantly. We conclude that, in our
simulation of the dripline, the realistic NN correlation is not
enough to generate a dineutron halo.
\begin{figure}[b!]
\vspace{-1cm}
\centerline{
    \parbox[t]{0.45\textwidth}{
\centerline{\psfig{figure=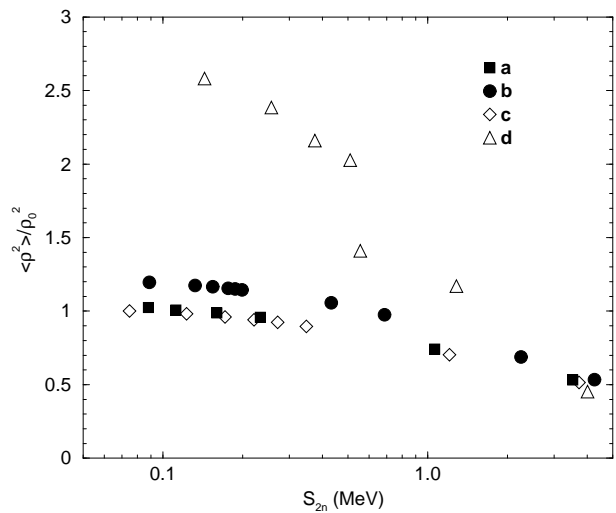,width=0.5\textwidth}}
\vspace{-1cm}
    \caption{The ratio of the rms hyper-radius and the
    scaling length for the three body system based on
    $^{12}$Be: a) including both the NN interaction
    and core deformation; b) including only core deformation;
    c) including only the NN interaction; d) switching both
    the NN interaction and the core deformation to zero. }
\label{fig1a}}
}
\end{figure}

Information identical to Fig. \ref{fig1} could be expressed
through the evolution of
the occupation of the $s_{1/2}^2$ orbital near threshold
(where both $l_x=0$ and $l_y=0$).
Here again, we find that the NN interaction
reduces somewhat the occupancy of the orbital, and this reduction
is accentuated when including the core deformation.
Essentially, core deformation mixes in
many other components of higher angular momentum,
reducing the occupation of the s-wave orbitals.

In \cite{dima} a sufficient condition for a halo is given
in term of the rms of the hyper-radius. One needs to define 
a typical hyper-radius scale 
associated with the two body forbidden regions, $\rho_0$
as in Eq.(7) of \cite{dima}. 
In the $^{12}$Be example, the relevant quatities are the radius 
for the $^{10}$Be-n interaction and the scattering length for the nn
interaction. Then the scale becomes  $\rho^2_0  =  53$ fm$^2$. 
The condition for a halo is now $<\rho^2>/\rho_0^2>2$.
In Fig. \ref{fig1a} we plot this ratio as a function of
binding energy for the four cases presented above.
A typical halo develops naturally when there are no
core-n or nn correlations.
It is clearly seen that both core excitation/deformation
and the NN interaction hinder the development of the halo.


\subsection{Other ways of reaching the dripline}

There are several issues associated with this simplified
prescription of reaching the dripline. Probably, nature is not
kind enough to preserve the same parameters in Eq.
(\ref{eq:potential}), when moving away from the valley of
stability. For instance, one expects the deformation of the core
to vary throughout the nuclear chart, when adding neutrons to
reach the corresponding dripline nucleus. For this reason, we have
repeated the above calculations, starting with the same n-core
interaction, but varying the binding energy now through the
deformation parameter  $\beta_2 = 0.05 \rightarrow 1$ and found
that the qualitative features discussed in the previous section
remained unchanged.

Moreover, we have explored a combination of possible initial
compositions of the subsystem '$^{11}$Be=$^{10}$Be+n'. We have
tried other geometries for the n-core interactions ($R_{ws}$ and
$a_{ws}$) which produce a starting '$^{12}$Be' with different
ground state dominant components, within the same model space.
We also tried variations on the deformation parameter. 
Nowhere in the explored
parameter space did we find the possibility of a two-neutron halo
developing with the inclusion of NN and n-core correlations.

\subsection{Heavier systems}

In the deformed nuclear region of intermediate mass, the competing
shells are no longer $2s_{1/2}$, $1d_{5/2}$ and $1p_{1/2}$. One
possibility among many is the example explored in \cite{pair1}
where $3s_{1/2}$, $1g_{7/2}$ and $2d_{5/2}$ play a role. Ideally
we would have repeated the calculations for the $3s_{1/2}$ and
neighboring shells. The number of forbidden states as well as the
number of channels in our model space would drastically increase.
Then, the calculations with core excitation would no longer be
feasible. Nevertheless, the larger partial waves, if anything,
will only enhance the findings for the case discussed above. It is
the recoil of the heavy core that will be smaller than for A=10,
which may reduce the impact of the core couplings. Also the
typical deformation parameter is smaller than $\beta_2=0.67$. 
We have thus repeated the calculations by artificially increasing the
mass of the core to A=100 whilst keeping the same shell structure.
We take this estimate to be an upper limit for the existence of a halo
in this region.
\begin{figure}[t!]
\vspace{-1cm}
\centerline{
    \parbox[t]{0.45\textwidth}{
\centerline{\psfig{figure=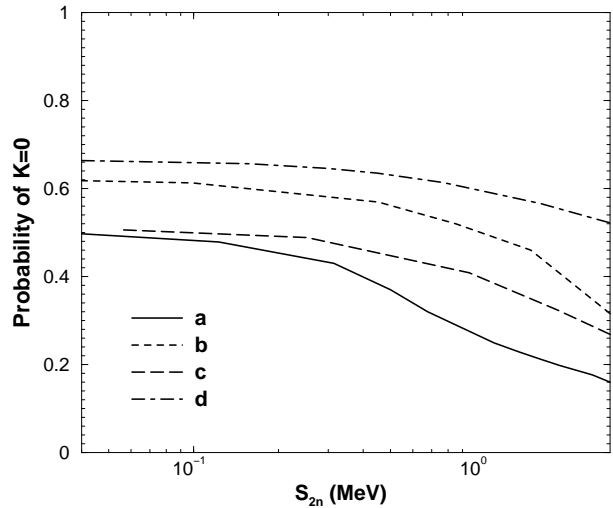,width=0.5\textwidth}}
\vspace{-1cm}
    \caption{Probability of the K=0 component in the
    ground state of a three-body core+n+n system based on
    a A=100 core: a) including both the NN interaction
    and core deformation ($\beta_2=0.3$); 
    b) including only core deformation ($\beta_2=0.3$);
    c) including only the NN interaction; d) switching both
    the NN interaction and the core deformation to zero. }
\label{fig2}}
}
\end{figure}

We refit the n-core interaction to produce a ground state at
around $S_{2n}=4$ MeV, the $1/2^+$ and $1/2^-$ level ordering for
the n-core subsystem, using $\beta_2=0.3$, more adequate for heavier
systems. The resulting ground state wavefunction has a
configuration similar to $^{12}$Be; mainly an admixture of
$2s_{1/2}$,  $1d_{5/2}$ and $1p_{1/2}$ with a significant amount
of core excitation. We then reduce the overall strength of the
n-core interaction to simulate the proximity of the dripline. The
separate effects of the NN interaction and core deformation are
shown in Fig. \ref{fig2}. Expectedly the role of deformation
is reduced compared to section 3.1. The effect of the NN
interaction alone is now larger than the effect of deformation.
Interestingly, the calculation including both, the
NN interaction and core deformation/excitation, 
is at the border line $\lim_{S_{2n}\rightarrow 0}P(K=0) \approx 0.5$.

We also checked whether the NN interaction is sufficient to form a
'dineutron'+core system as then one may
obtain a halo system in the two body sense. The dineutron
component with s-motion relative to the core, does increase to
$68$\% but throughout the simulated path toward the dripline
the average distance between the two valence neutrons 
is increasing and is always larger than the average distance
between the neutrons and the core. Consequently a dineutron
picture does not make sense.

\section{Conclusions}

We have explored the configuration of the ground state
wavefunction of the three-body nuclear system when approaching
threshold. The aim of the work was to determine whether core
deformation and/or pairing of the valence nucleons would hinder
the appearance of the two-neutron halo phenomenon. The three-body
model with core excitation is most adequate to explore these
physical aspects explicitly. A variety of three-body calculations
based on the $^{12}$Be model with core excitation were performed.
Our results show that both the NN tensor interaction, which goes
beyond the usual pairing in HFB, and the couplings due to core
deformation can significantly reduce the probability of a three
body halo developing when approaching the neutron dripline. 

For this work, the rotor core model includes a $0^+$ core g.s. and
its first excited state $2^+$. The addition of the $2^+$ excited
state has brought into the picture couplings to higher angular
momentum: namely, the core-n system in its ground state contains
not only the s-wave with the core in the g.s. but also a d-wave
with the core in the $2^+$ state. This is an essential feature for
the disappearance of the halo phenomenon. Of course there may be
situations where coupling to higher angular momentum is obtained
just with reorientation effects or, on the other end, there may be
particular cases where no additional angular momentum is added to
the system even including the most important excited states of the
core. However, in the general case, the core-n single particle
s-state, which would develop into a halo state at threshold, will
couple to higher angular momentum when including collective
degrees of freedom of the core, hindering the appearance of the halo.

Our conclusions are based on the assumption of the three-body model
with core deformation/excitation. This model, while of interest
for qualitative features, is certainly not appropriate for
quantitative predictions. Improvements on the description
of the core can be obtained within AMD \cite{kimura}.
In order to predict the existence
or non-existence of a heavy halo, one needs a step further:
a  fully antisymmetric self-consistent mean field model that
includes both pairing and deformation to all orders \cite{hfb}.
In the near future there is a plan to explore the dripline around
the deformed mass region, using the fully self-consistent
microscopic mean field which includes both static pairing and
quadrupole deformation (e.g. \cite{duguet}).

\vspace{0.5cm}
{\bf Acknowledgements}

We are grately in debt to Thomas Duguet for many discussions
related to mean field theories. We thank Ian Thompson and
Dmitri Fedorov for useful comments to the manuscript.

\end{document}